# Replica symmetry breaking in specially designed TiO₂ nanoparticles-based dye-colloidal random laser

Pablo I. R. Pincheira,[1] Andréa F. Silva,[2] Sandra J. M. Carreño,[1] Serge I. Fewo,[3] André L. Moura,[1,4,*] Ernesto P. Raposo,[5] Anderson S. L. Gomes[1] and Cid B. de Araújo[1]

[1]*Departamento de Física, Universidade Federal de Pernambuco, 50670-901, Recife-PE, Brazil*
[2]*Programa de Pós-Graduação em Ciências de Materiais, Universidade Federal de Pernambuco, Recife 50670-901, Brazil*
[3]*Laboratory of Mechanics, Department of Physics, University of Yaoundé I, Cameroon*
[4]*Grupo de Física da Matéria Condensada, Núcleo de Ciências Exatas – NCEx, Campus Arapiraca, Universidade Federal de Alagoas, 57309-005, Arapiraca-AL, Brazil.*
[5]*Laboratório de Física Teórica e Computacional, Departamento de Física, Universidade Federal de Pernambuco, 50670-901, Recife-PE, Brazil*

*\*Corresponding author. E-mail:* andre.moura@fis.ufal.br

**ABSTRACT**

By using specially designed nanoparticles scatterers which prevents photodegradation of the dye, we present clear demonstration of replica symmetry break and a photonic paramagnetic to spin-glass phase transition in a random laser operating in the incoherent feedback regime based on ethanol solution of Rhodamine 6G and amorphous $TiO_2$ nanoparticles.

PACS number(s): 42.55.Zz, 42.55.Mv, 75.50.Lk.

# I. INTRODUCTION

Replica symmetry breaking (RSB) is a concept inherent to the theory of spin glasses and complex systems [1], which describes how identical systems prepared under identical initial conditions can reach different states. RSB was predicted [2,3] and demonstrated [4], for the first time in any physical system, using solid state based random lasers (RLs) [4,5]. RLs are cavity-less lasers based on disordered systems with the feedback for laser action provided by the multiple scattering of light [6-8], whose basic properties and applications have been reviewed by several authors [9-11]. Mode competition in RLs has attracted large attention [12,13], and they have been used as a platform to study disordered complex systems. By investigating the distribution of correlations between RL intensity fluctuations from pulse-to-pulse, Ghofraniha et al. [4] defined an analogue function to the Parisi overlap order parameter ($q_{max}$) and found evidence of a transition from the photonic paramagnetic to a glassy phase of light – a photonic spin-glass phase [2,3,5].

As described in ref. [4], RSB was manifested in a in which resonant modes are revealed by multiple spikes in the RL spectrum, when high resolution spectral measurements were performed. For lower spectral resolution, the spikes are not seen in the spectral data [4]. However, attempts to observe RSB in incoherent feedback RLs, in which the RL spectrum is initially narrowed due to amplified spontaneous emission (ASE) and further narrowing occurs above the RL threshold, were not successful, as also reported in ref. [4]. Although still a matter of debate, these two regimes where spikes with linewidths typically below 1 nm appear, are interpreted as due to interference among the paths undergone by the photons, and in this case it is known as coherent (or resonant) feedback. These paths would arise from closed loops [9]. Conversely, spectral signature typical of those observed in colloidal based RL, in which many overlapping modes lead to a smooth but spykeless spectrum, has been denominated as incoherent (or nonresonant) feedback. [9]. More recently, extension of RSB to RLs operating in the incoherent feedback regime was demonstrated in $Nd^{3+}$ doped nanocrystal powders, where the spectral RL signature is a smooth profile [14].

The contrasting results between the $Nd^{3+}$ and dye-based RLs are not intuitively expected and a deeper investigation of the dye-based RLs behavior was the motivation for the present work. However, the most important point of our work is to provide a colloidal system that can be used for a timeframe long enough, without altering the RL properties, such that the system stays the same from shot to shot, and the disorder configuration is the same. As will be shown, in this case RBS was readily observed.

Dye-based RLs are highly efficient, and have been used as a platform for demonstration of optical effects such as the bichromatic emission due to monomers and dimers in Rhodamines [15-17], multi-photon excitation [18-19], and low spatial coherence [20]. In spite of the promising characteristics of dye-based RLs, the dye photodegradation, enhanced due to the presence of the scatterers is a drawback for the basic investigation and practical application of these systems [21,23]. The most used scatterer particles material is $TiO_2$ in the rutile phase due to its high refractive index ($\sim$ 2.6 at 632.8 nm) that implies a large index contrast with the dye solution. However, the energy bandgap of rutile is $\sim$ 3.2 eV, which allows, by exciting with visible light, two-photon excitation of electrons from the valence to the conduction band. Then, a photocatalytic process is started leading to fast dye degradation [24]. Attempts to minimize the photodegradation were made using scatterer particles based on materials with large energy bandgap [21-23]. Indeed, silica particles, which have energy bandgap of $\sim$ 5.0 eV and refractive index of $\sim$ 1.7 in the visible range, have been used and lower photodegradation was observed [21-23]. However, two other important points neglected by most of researchers are the particles precipitation and the chemical bonding between the particles and the cuvette walls that contribute to reduce the efficiency and lifetime of the RLs. Another important consequence is the impossibility of making long time observation of the RL statistics such as those in solid state RL [4,14].

The system studied in this work consisted of amorphous $TiO_2$ nanoparticles suspended in an ethanol solution containing Rhodamine 6G. However, differently from the majority of reports in the literature, the particles were functionalized to minimize precipitation in the time scale of the experiments, and do not make chemical bonds with the silica at the cuvette walls. As a consequence the dye photodegradation was largely reduced and a low RL threshold was obtained. On the other side, the very slow photodegradation allowed measurements of the intensity fluctuations statistics using stable samples, without degradation, for more than subsequent $10^5$ laser shots. Therefore, correlations between intensities from shot-to-shot could be analyzed and our findings show that RSB is unveiled mainly near the threshold. Furthermore, although the narrowing of the emitted spectra is smooth, the Parisi overlap parameter reveals a first order phase transition that coincides with the energy threshold of the RL. These results show that the conclusions in refs. [4,5] about the use of colloidal RLs, particularly with $TiO_2$, should be attributed to the characteristics of the colloid sample used, and can be overcome. It is not a fundamental property of RLs. Once the dye solution with scatterer particles is carefully prepared, the RSB behavior is similar to that of a solid state based RL, as in ref. [4].

## II. EXPERIMENTAL DETAILS

Optical experiments were conducted in a similar scheme described in recent publications [14,25,26] where details are presented. The excitation source was the second-harmonic at 532 nm of a Nd-YAG laser operating at 5 Hz with beam area of 0.7 mm$^2$, incidence angle of 30° with respect to the normal to the surface, and the excitation pulse energy (EPE) varied up to 4.60 mJ. For each excitation pulse, a single spectrum was recorded, and a sequence of 1000 spectra was collected for each EPE.

The RL samples consisted of ethanol solution of Rhodamine 6G and $TiO_2$ nanoparticles at concentrations of $10^{-4}$ M and $6.7 \times 10^{11}$ cm$^{-3}$, respectively, placed in a quartz cuvette with dimensions $10 \times 10 \times 50$ mm$^3$.

Commercial $TiO_2$ particles, acquired from Dupont Inc, with the crystalline structure of rutile (average diameter of 250 nm) were initially tested. As shown in the text these particles present fast precipitation and attach to the silica cuvette walls. In order to compensate for these deleterious aspects, amorphous $TiO_2$ nanoparticles with average diameter of 168 nm were synthesized by the sol-gel method as described in ref. [27]. Firstly, ethanol P.A. (EtOH – 99.5%, Sigma Aldrich) and acetonitrile P.A. (AN – 99.8%, Sigma Aldrich) were mixed in a volume ratio of 1:1, EtOH/AN, obtaining a volume of 5 mL. Afterwards, 0.1 M (170 μL) of titanium tetrabutoxide (TBO, 97% Sigma Aldrich) was introduced in the solution under nitrogen atmosphere. The final solution is labeled as solution A. Another mixture of EtOH/AN (1:1) at 5 mL, solution B, was prepared and, after that, 0.2 M of $NH_4OH$ P.A. (40 μL) and 1.0 M of Milli-Q water (90 μL) were added to the solution. To obtain the $TiO_2$ nanoparticles, solution B was added to solution A under magnetic stirring at 550 rpm for 5 minutes. To stop the reaction, ethanol at a volume of 10 mL was added to the solution. Subsequently, washing of the colloid was made at 10,000 rcf for 5 minutes and the supernatant was extracted. Then, resuspension in ethanol was performed in an ultrasound bath. This process was repeated 5 times and then, after the fifth resuspension, the $TiO_2$ nanoparticles were suspended in 5 mL of ethanol.

The amorphous $TiO_2$ nanoparticles do not precipitate in the ethanol solution because they have hydroxyl groups in their surface that make hydrogen bonding with the ethanol molecules. Moreover, due to the hydrogen bonding, the $TiO_2$ particles do not attach to the silica walls of the cuvette.

### III. RANDOM LASER CHARACTERIZATION

For the synthesized $TiO_2$ particles, the x-ray diffraction pattern and the particles' size distribution are shown in Fig. 1. The absence of sharp diffraction peaks is an indication that 100% of the nanoparticles are amorphous.

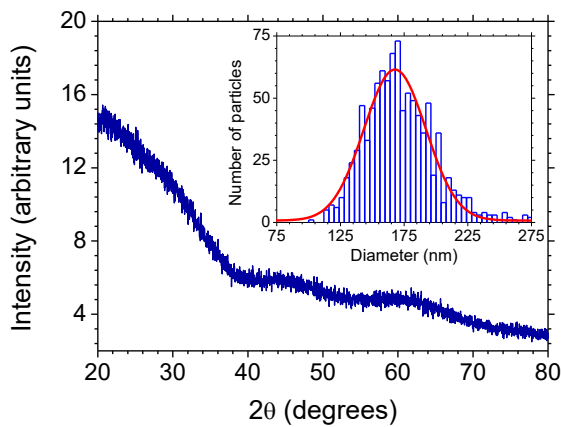

FIG. 1. (Color online) Functionalized $TiO_2$ nanoparticles characterization. X-ray diffraction pattern of the synthesized $TiO_2$ particles. The inset shows the particle size distribution.

Characterization of the RL emission was performed, as usually reported in the literature, by recording the emitted spectra for various EPE. Figure 2(a) shows the emitted spectra for EPE smaller (0.015 mJ), around (0.12 mJ) and larger (4.60 mJ) than the RL threshold of 0.11 mJ, from where the bandwidth narrowing and redshift of the spectra are evident. Figure 2(b) presents the bandwidth narrowing, characterized by the full width at half maximum (FWHM), and the peak intensity as a function of the EPE. The transition from amplified spontaneous to RL emission is smooth. The solid line shown is a sigmoidal fit to the FWHM data for better determination of the EPE threshold (at 0.11 mJ) defined as in ref. [8] as the inflection point of the curve.

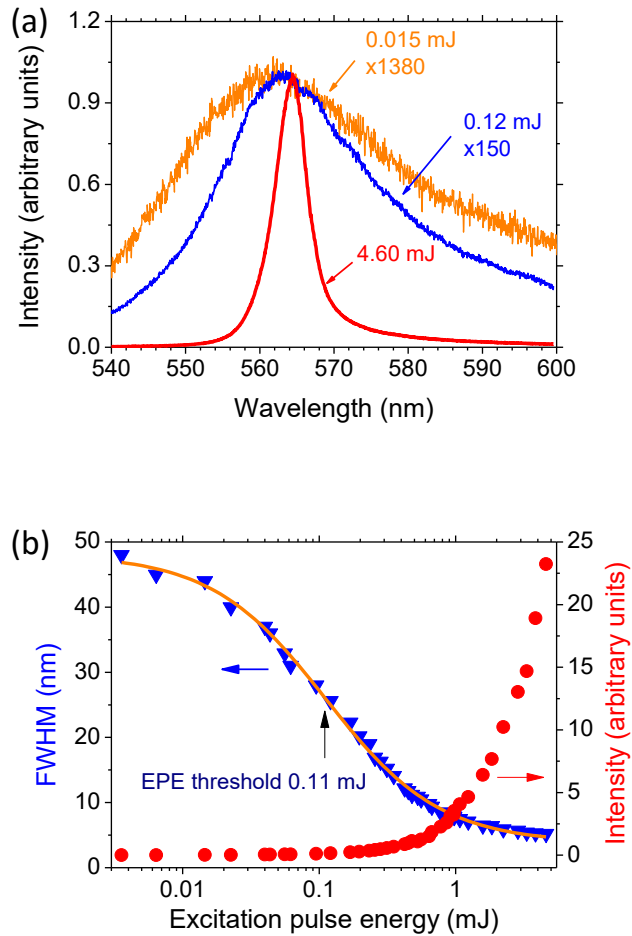

FIG. 2. (Color online) Random laser characterization for a sample with amorphous $TiO_2$ nanoparticles. (a) Emitted spectra for excitation pulse energies of 0.015, 0.12, and 4.60 mJ,

which are smaller, around, and larger than the random laser threshold of 0.11 mJ. (b) Full width at half maximum (FWHM) and maximum intensity of the emitted spectra for different excitation pulse energies. The solid line is a sigmoidal fit to the FWHM data.

Visual comparison of samples with commercial (non-functionalized) and the synthesized $TiO_2$ nanoparticles at the same concentration can be appreciated in the inset of Fig. 3. The color of the solutions looks different from one sample to another because after some time the commercial $TiO_2$ particles attach to the cuvette walls preventing light transmission, as can be inferred by viewing the upper part of the flasks. It is clear that the bonding of $TiO_2$ to the silica at the flask walls is well-pronounced for commercial $TiO_2$. Also, precipitation of the particles is noted by accumulation of powder at the bottom of the cuvette.

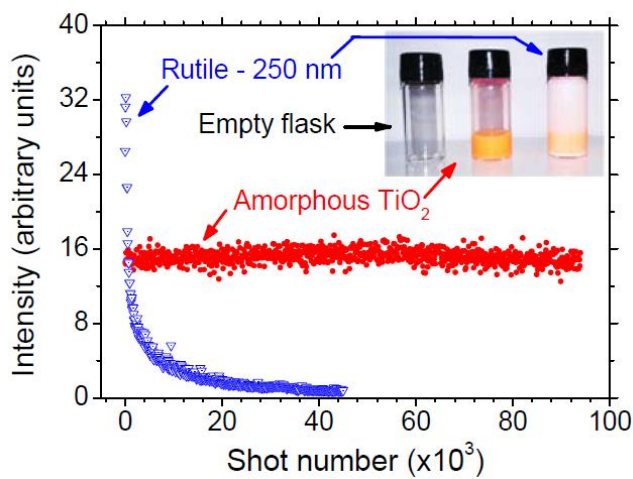

FIG. 3. (Color online) Evaluation of dye photodegradation. Random laser peak intensity as a function of the number of shots with the measurements performed at 5 Hz and excitation pulse energy of 4.00 mJ for solutions of ethanol, Rhodamine 6G and amorphous $TiO_2$, and rutile. The inset shows, from right to left, the solutions for rutile, amorphous $TiO_2$, and an empty flask.

Dye photodegradation analysis was done by exciting the samples with EPE of 4.00 mJ over more than 40,000 shots. Figure 3 shows the RL intensity as a function of the number of shots for both solutions. Unlike commercial $TiO_2$, the in-house synthesized $TiO_2$ does not present any indication of photodegradation for at least $10^5$ shots (at 5 Hz), being therefore useful to RLs studies and applications that require long exposure to the incident optical pulses.

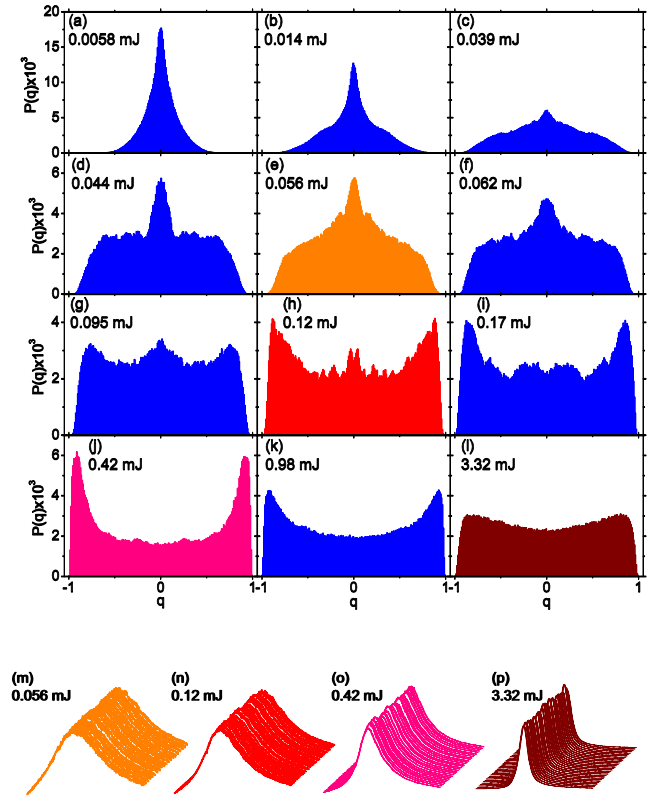

FIG. 4. (Color online) Open cavity laser (random laser) overlap distribution functions. Excitation pulse energies: (a) 0.0058, (b) 0.014, (c) 0.039, (d) 0.044, (e) 0.056, (f) 0.062, (g) 0.095, (h) 0.12, (i) 0.17, (j) 0.42, (k) 0.98, and (l) 3.32 mJ. (m) to (p) show the pulse-to-pulse fluctuations in the emitted spectra for excitation pulse energies of 0.056, 0.12, 0.31, and 4.32 mJ, respectively.

## IV. REPLICA SYMMETRY BREAKING

RSB is determined as in ref. [4] by evaluating the correlation function that measures pulse-to-pulse fluctuations in the spectral intensity averaged over $N_s$ laser shots, using the expression

$$q_{\gamma\beta} = \frac{\sum_k \Delta_\gamma(k)\Delta_\beta(k)}{\sqrt{\left[\sum_k \Delta_\gamma^2(k)\right]\left[\sum_k \Delta_\beta^2(k)\right]}},$$

where $\gamma$ and $\beta = 1, 2, \ldots, N_s$, with $N_s = 1000$, denote the pulse (replica) labels. The average intensity at the wavelength indexed by $k$ reads $\bar{I}(k) = \sum_{\gamma=1}^{N_s} I_\gamma(k) / N_s$, and the intensity fluctuation is measured by $\Delta_\gamma(k) = I_\gamma(k) - \bar{I}(k)$. Each laser shot is considered a replica, i.e. a copy of the RL system under identical experimental conditions. The probability density function (PDF), $P(q)$, describes the distribution of replica overlaps $q = q_{\gamma\beta}$, signalizing a photonic paramagnetic or a RSB spin-glass phase whether it peaks at $q = 0$ (no RSB) or at values $|q| \neq 0$ (RSB), respectively.

Since the RL does not photodegrade, the initial conditions are kept the same when comparing one spectrum to the other for each EPE value. The overlap distribution function is represented in Fig. 4 for the EPE values indicated. Also, the pulse-to-pulse intensity fluctuation over 20 spectra is represented in Figs. 4(m) to (p) for EPE of 0.056, 0.12, 0.31, and 4.32 mJ, respectively. It is worth reminding here that, in the RL framework, the energy plays the role of the inverse temperature in spin glass theory of disordered magnetic systems [4].

The order parameter $q_{max}$ is represented in Fig. 5 together with the full width at half maximum (FWHM) of the emission bandwidth as a function of the EPE, in which a transition from the photonic paramagnetic to the spin-glass phase is clearly seen. Although the FWHM changes smoothly as the EPE increases, a first order transition is observed by the abrupt change of $q_{max}$ around the EPE value of 0.11 mJ, coinciding with the EPE threshold determined from the FWHM variation. We also remark that the spectral narrowing for EPE smaller than the RL threshold ($< 0.11$ mJ) in Fig. 5 is attributed to the ASE. However, by looking at the $q_{max}$ parameter value, the conclusion is that the system is still in the photonic paramagnetic regime ($q_{max} = 0$), and the modes (analogous to the spins) totally uncorrelated. Just above the threshold, while the spectral narrowing is still taking place, the system is already in the photonics spin glass regime, and the RSB signature appears due to the mode correlation or anti-correlation. The RSB occurs in the transition from ASE to laser regime and beyond.

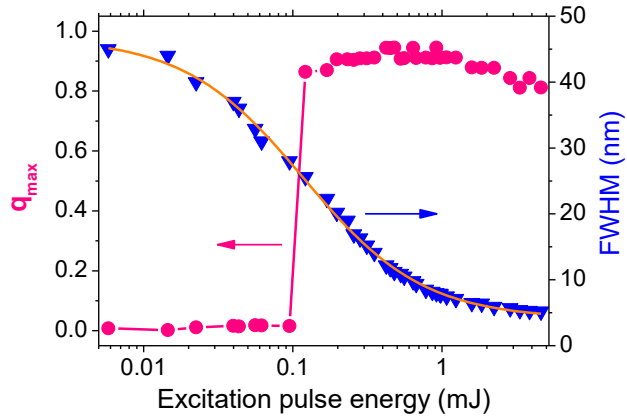

FIG. 5. (Color online) Agreement between photonic paramagnetic to spin-glass phase transition and the random laser threshold. Parisi overlap parameter and bandwidth dependence with the excitation pulse energy (in logarithm scale).

As a final remark, we emphasize that the Parisi overlap parameter can be used as an indicator of the threshold for laser action in random media. As discussed above, the photonic phase transition coincides with the laser threshold, as observed in other studied solid state based RLs [4,14].

## V. CONCLUSIONS

In summary, amorphous $TiO_2$ nanoparticles were synthesized and exploited as scatterers for random lasers (RLs) characterization and application. Owing to hydroxyl groups on the $TiO_2$ surface, the nanoparticles are stable in ethanol solution and do not make chemical bonds with the silica at the cuvette walls. In solution with Rhodamine 6G, it provided a non-photodegraded and efficient RL. We used this system as a platform for the first demonstration of replica symmetry breaking in an open cavity dye-based random laser operating in the incoherent feedback regime. Although the spectral narrowing is smooth as the excitation pulse energy increases, a first-order phase transition, evaluated by the Parisi overlap parameter, from the paramagnetic to the spin-glass phase was observed. The herein presented results demonstrate that the intensity fluctuations behavior of colloidal based RLs is similar to the behavior previously reported for solid state RLs regarding their analogy with spin glass magnetic systems.

## ACKNOWLEDGEMENTS

We acknowledge the financial support from the Brazilian Agencies: Conselho Nacional de Desenvolvimento Científico e Tecnológico (CNPq), Coordenação de Aperfeiçoamento de Pessoal de Nível Superior (CAPES), and Fundação de Amparo à Ciência e Tecnologia do Estado de Pernambuco (FACEPE). The work was performed in the framework of the National Institute of Photonics (INCT de Fotônica) and PRONEX-CNPq/FACEPE projects. A.F.S. acknowledges CNPq for a scholarship. P.I.P.R. and S.J.M.C. acknowledges CAPES for their scholarship. A.L.M. acknowledges CNPq for a postdoctoral fellowship.